\documentclass{aa}
\usepackage{epsfig}

\def\etal{{\rm et al.\thinspace}}
\def\eg{{\rm e.g.\ }}

\def\ie{{\rm i.e.\ }}
\def\cf{{\rm cf.\ }}

\def\spose#1{\hbox to 0pt{#1\hss}}
\def\ltsimm{\mathrel{\spose{\lower 3pt\hbox{$\sim$}}
	\raise 2.0pt\hbox{$<$}}}
\def\gtsimm{\mathrel{\spose{\lower 3pt\hbox{$\sim$}}
	\raise 2.0pt\hbox{$>$}}}
\def\Mdot{\hbox{${\dot M}$}}

\def\km{{\rm\thinspace km}}
\def\cm{{\rm\thinspace cm}}
\def\s{{\rm\thinspace s}}
\def\yr{{\rm\thinspace yr}}
\def\kmps{\hbox{${\rm\km\s^{-1}\,}$}}

\def\Msol{\hbox{${\rm\thinspace M_{\odot}}$}}
\def\Msolpyr{\hbox{${\rm\Msol\yr^{-1}\,}$}}

\begin{document}
   
\title{The dominant X-ray wind in massive star binaries}

\author{J.M. Pittard\inst{1} \and I.R.Stevens\inst{2}}

\institute{Department of Physics \& Astronomy, The University of Leeds, 
        Woodhouse Lane, Leeds, LS2 9JT, UK
 \and Department of Physics \& Astronomy, The University of Birmingham,
        Edgbaston, Birmingham, B15 2TT, UK}

\offprints{J. M. Pittard, \newline
\email{jmp@ast.leeds.ac.uk}}

\date{Received <date> / Accepted <date>}

\abstract{We investigate which shocked wind is responsible for the 
majority of the X-ray emission in colliding wind binaries, an issue 
where there is some confusion in the literature, and which we show is 
more complicated than has been assumed. We find that where both winds
rapidly cool (typically close binaries), the ratio of the wind speeds
is often more important than the momentum ratio, because it controls the
energy flux ratio, and the {\em faster} wind is generally the dominant emitter.
When both winds are largely adiabatic (typically long-period binaries), the
slower and denser wind will cool faster and the {\em stronger} wind generally
dominates the X-ray luminosity.
}

\titlerunning{The dominant X-ray wind in massive star binaries}
\authorrunning{Pittard \& Stevens}

\maketitle

\label{firstpage}

\section{Introduction}
\label{sec:intro}
The violent wind-wind collision in massive star binaries creates a region 
of high temperature shock-heated plasma, which can contribute to the total 
system emission at radio, infrared, optical, ultraviolet and X-ray 
wavelengths. Over the last 20 years, theoretical models have focussed 
mainly on the dynamics of the stellar winds and the wind collision 
zone (WCZ), and on the resulting X-ray emission (Pittard \cite{P2000} 
and references therein). As the X-ray emission is dependent on 
the physical conditions within the WCZ and on the distribution and 
properties of the unshocked attenuating wind material, observations provide 
information on basic parameters of the system (see Stevens \etal 
\cite{S1996}; Zhekov \& Skinner \cite{ZS2000}; Pittard \& Corcoran
\cite{PC2002}).
 
To interpret such observations, it is useful to know which wind dominates 
the X-ray emission. For instance, if there is a velocity difference 
between the winds, we would like to be able to predict how hot the broad-band 
emission will be, or if there are abundance differences, how strong or 
weak certain lines will be (this is of particular relevance now that
grating observations can resolve line profiles). 

Simple expressions for the X-ray luminosity from each of the shocked winds 
were presented in a complex, yet elegant paper, which provides some answers
to these questions in terms of fundamental parameters of the system 
(Usov \cite{U1992}). Their applicability has resulted in their common use 
in the literature and in observing proposals. However, we have recently 
discovered inconsistencies between results from these expressions and those 
determined from numerical models, leading to some confusion on the issue of 
the dominant X-ray emitting wind. In this letter we reinvestigate previous
conclusions in the literature.

\section{Estimates of $L_{1}/L_{2}$}
\label{sec:previous}

Let us define $L_{1}$ as the X-ray luminosity from the shocked  
wind with the greater momentum flux (\ie the `stronger' wind), and $L_{2}$ 
the equivalent from the shocked weaker wind. 
Analytical estimates in the literature of the ratio of $L_{1}/L_{2}$ exist 
for two limiting cases: the radiative limit, where the cooling timescale 
for the hot gas from both winds is assumed small in comparison to the 
timescale for flow of this gas out of the system (\ie $\chi < 1$, see 
Sec.~\ref{sec:adiabatic_limit}); and the 
adiabatic limit, where the opposite is true. We will first focus on these 
limiting cases, before discussing the behaviour of $L_{1}/L_{2}$ between 
these limits.

\subsection{$L_{1}/L_{2}$ in the radiative limit}
\label{sec:rad_limit}

In this limit, the entire kinetic energy thermalized by the shocks is
immediately radiated (normally with the majority at X-ray energies), and the 
region of shocked gas is thin. Binary systems with a small orbital separation
are favoured, and examples near this limit include V444~Cyg 
(WN5 + O6; P = 4.2~d; Corcoran \etal \cite{C1996}) and DH~Cep 
(O5.5-6V + O6.5-7V; P = 2.11~d; Penny \etal \cite{PGB1997}).

\begin{figure}
\begin{center}
\psfig{figure=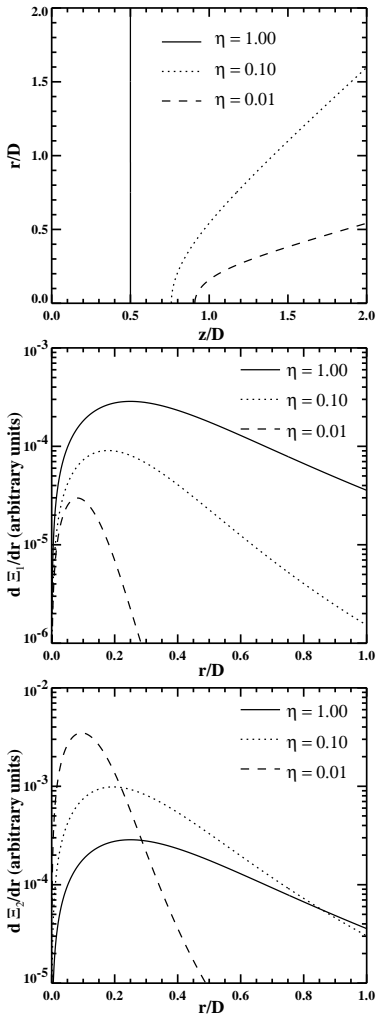,width=5.0cm}
\end{center}
\caption[]{Results for a wind-wind collision in the radiative limit, with 
equal and spatially invariant pre-shock speeds. The top panel shows the 
position of the contact discontinuity as a function of wind momentum ratio, 
$\eta$. Star 1 is located at (r,z) = (0,0), and star 2
at (r,z) = (0,1). The middle and bottom panels show the value of 
d$\Xi$/d$r$ (see text) as a function of $r$ and $\eta$ for the wind of star 
1 and star 2 respectively. Note that the majority of the thermalized 
kinetic power of the winds occurs well within $r = D$ (see also 
Table~\ref{tab:rad_lim2}).}
\label{fig:rad}
\end{figure}

Let us first consider the region of the WCZ which lies directly 
between the stars. As there is a momentum balance, the relative
kinetic energy fluxes are proportional to the ratio of the wind speeds, so 
emission from this volume should be dominated by gas from the star with the 
faster wind. If, for simplicity, we initially assume that the pre-shock 
wind speeds are spatially invariant, we find that the majority of the
thermalized energy in the WCZ occurs close to the line of centres (see 
Fig.~\ref{fig:rad}). Under these conditions, the faster wind 
dominates the {\em total} emission.

If the pre-shock wind speeds are approximately equal, neither star
should significantly dominate the total emission. In such circumstances, 
one might expect secondary considerations, such as the wind momentum ratio, 
to become important.  
Luo \etal (\cite{LMM1990}) noted that if $v_{1} \approx v_{2}$, the weaker
wind should be more luminous, based on the premise that it
has, on average, a larger velocity component normal to the shock front:
referring to V444~Cyg they argue that, ``Because of this, the shocked O6 
wind dominates the X-ray emission from the shocked region''. However, we show 
below that even for very large wind momentum ratios, the stronger wind still 
accounts for almost half of the total emission.
Therefore Luo \etal's use of the word ``dominates'' goes too far, although 
their basic conclusion that the O6 wind is the majority emitter is consistent
with our analysis. Complicating factors, such as wind acceleration, are 
discussed later.

We can obtain analytical estimates for $L_{1}/L_{2}$ from the
relevant equations in Usov (\cite{U1992}). For the radiative limit, 
the total X-ray luminosity from the external
wind shock is (\cf his Eq.~88) $L_{1} = 2.3 \; \rho_{\infty} \; 
v_{1}^{3} \; z_{\rm OB}^{2}$,
where $\rho_{\infty} = \Mdot_{1}/4 \pi D^{2} v_{1}$,
$z_{\rm OB}$ is the distance of the contact discontinuity from the centre of
the star with the weaker wind ($= \sqrt{\eta} D/(1 + \sqrt{\eta})$), 
$D$ is the stellar separation, and $\eta$ is the momentum ratio of the
winds ($= \Mdot_{2} v_{2}/\Mdot_{1} v_{1}$). Usov (\cite{U1992}) 
parameterizes the luminosity from the internal wind shock as 
$L_{2} = 0.22 \Mdot_{2} v_{2}^{2}$.

With the same parameter values as used by Luo \etal 
(\cite{LMM1990}) for V444~Cyg, these equations yield $L_{1}/L_{2} = 0.59$
(\ie that the weaker wind is the dominant X-ray emitter),
in agreement with the statement in Luo \etal (\cite{LMM1990}).
From Usov (\cite{U1992}) we find that $L_{1}/L_{2} \propto 
\Mdot_{1}v_{1}^2/\Mdot_{2}v_{2}^2
\approx \rho_{1}v_{1}^3/\rho_{2}v_{2}^3$,
which at the stagnation point is proportional to $v_{1}/v_{2}$, 
such that as expected the faster wind will normally be the dominant X-ray 
emitter. While this agreement is satisfying, an indication of potential 
problems with the equations in Usov (\cite{U1992}) 
is revealed by considering the predicted value of $L_{1}/L_{2}$ for 
identical winds: in this case we obtain $L_{1}/L_{2} = 0.2$, instead of unity!

We have therefore performed a numerical calculation of the kinetic energy 
flux normal to the wind shocks to precisely determine $L_{1}/L_{2}$ for 
various wind parameters. The position of the contact discontinuity 
(which is also the position of the wind shocks since the wind collision 
region is thin) was obtained from integration of Eq.~4 in Stevens \etal 
(\cite{SBP1992}). We find that the value of $L_{1}/L_{2}$ primarily depends on
the ratio of the wind speeds and to a much lesser extent on the wind momentum
ratio. Results are shown in Fig.~\ref{fig:rad} and in Table~\ref{tab:rad_lim},
where we have defined $\Xi$ to be the fractional wind 
kinetic power normal to the contact discontinuity 
\ie $L_{1} = 0.5\Xi_{1}\Mdot_{1}v_{1}^{2}$ for the stronger wind. 

For $\eta = (0.01,\; 0.1,\; 1.0)$, we find that
$\Xi_{1} = (0.0042,\;0.033,\;0.167)$ and $\Xi_{2} = (0.564,\;0.403,\;0.167)$
respectively (the latter values being the analytical limit of 1/6), consistent
with the solid angle of the wind collision zone as viewed from the star with 
the stronger (weaker) wind decreasing (increasing) with
decreasing $\eta$. The ratio of Usov's equations yield values for 
$L_{1}/L_{2}$ which are too low in comparison to the exact 
numerical calculation by factors of (1.09, 1.74, 4.88)
for $\eta = (0.01,\; 0.1,\; 1)$ respectively, irrespective of the
ratio of $v_{1}/v_{2}$. Thus Usov's equations for the radiative limit are 
somewhat in error, being most accurate for low values of $\eta$. 

Fig.~\ref{fig:rad} also shows the gradient of $\Xi$ as a function of the 
off-axis distance, $r$. In Table~\ref{tab:rad_lim2} we
list the value of $r$ at which $\int_{r=0}^{r}({\rm d}\Xi/{\rm d}r) {\rm d}r$ 
is 50\% and 90\%
of the asymptotic value of $\Xi$. For equal strength winds, 50\% and 90\% of 
the maximum wind kinetic power thermalized by the shocks (and hence radiated) 
occurs within $r/D = 0.38$ and 0.95 (this off-axis distance is much smaller 
than the corresponding value for adiabatic winds \cf Luo \etal \cite{LMM1990}).
Therefore distortion of the WCZ by orbital motion should not significantly
alter our previous conclusions. We can also safely relax our assumption of
spatially invariant wind speeds as we only need to consider this 
ratio between the stars.

Finally, for close binaries we should also consider the potential role of
radiative acceleration, inhibition (Stevens \& Pollock \cite{SP1994}) and 
braking (Gayley \etal \cite{GOC1997}) of the winds on $L_{1}/L_{2}$. 
As already shown, the ratio of the wind speeds is the critical parameter 
in such systems, and it seems plausible to expect this to continue to be 
true in the presence of these 3 effects. In such binaries, it is probable
that neither wind will have room to reach terminal speed. Since the stronger 
wind will have more room to accelerate, we would normally expect it to be 
faster, and to dominate the emission. However, if there is substantial
radiative inhibition plus braking, this will cause the strong wind to be 
by far the slower and will likely shift the dominant energy generation back 
to the weaker wind. 

\begin{table}
\begin{center}
\caption{Ratio of the wind X-ray luminosities in the radiative limit assuming
spatially invariant pre-shock wind speeds. The value of $L_{1}/L_{2}$ 
primarily depends on the ratio of the wind speeds,
$v_{1}/v_{2}$, and to a lesser degree on the wind momentum ratio, $\eta$. 
Also tabulated are the ratios of the mass-loss rates and the wind kinetic 
power.}
\label{tab:rad_lim}
\begin{tabular}{ccccc}
\hline
$v_{1}/v_{2}$ & $\eta$ & $\Mdot_{1}/\Mdot_{2}$ & 
$\Mdot_{1} v_{1}^{2}/\Mdot_{2} v_{2}^{2}$ & $L_{1}/L_{2}$\\
\hline
2.000 & 10.0 & 0.05 & 0.200 & 2.43 \\
      & 1.00 & 0.50 & 2.000 & 2.00 \\
      & 0.10 & 5.00 & 20.00 & 1.64 \\
      & 0.01 & 50.0 & 200.0 & 1.48 \\
1.000 & 10.0 & 0.10 & 0.100 & 1.22 \\
      & 1.00 & 1.00 & 1.000 & 1.00 \\
      & 0.10 & 10.0 & 10.00 & 0.82 \\
      & 0.01 & 100. & 100.0 & 0.74 \\
0.500 & 10.0 & 0.20 & 0.050 & 0.61 \\
      & 1.00 & 2.00 & 0.500 & 0.50 \\
      & 0.10 & 20.0 & 5.000 & 0.41 \\
      & 0.01 & 200. & 50.00 & 0.37 \\
0.250 & 10.0 & 0.40 & 0.025 & 0.30 \\
      & 1.00 & 4.00 & 0.250 & 0.25 \\
      & 0.10 & 40.0 & 2.500 & 0.21 \\
      & 0.01 & 400. & 25.00 & 0.19 \\
\hline
\end{tabular}
\end{center}
\end{table}

\begin{table}
\begin{center}
\caption{Values of the off-axis distance, $r$, at which $\Xi_{1}$ and
$\Xi_{2}$ are 50\% and 90\% of their maximum value for various 
$\eta$, and assuming spatially invariant, equal speed winds.}
\label{tab:rad_lim2}
\begin{tabular}{ccccc}
\hline
$\eta$ & 0.5 $\Xi_{1,max}$ & 0.9 $\Xi_{1,max}$ & 0.5 $\Xi_{2,max}$ & 0.9 $\Xi_{2,max}$\\
\hline
1.00 & 0.38 & 0.95 & 0.38 & 0.95 \\
0.10 & 0.25 & 0.54 & 0.27 & 0.62 \\
0.01 & 0.10 & 0.19 & 0.11 & 0.22 \\
\hline
\end{tabular}
\end{center}
\end{table}

\subsection{$L_{1}/L_{2}$ in the adiabatic limit}
\label{sec:adiabatic_limit}

In direct contrast to the radiative limit, colliding wind systems are most 
likely to be near the adiabatic limit if the stellar separation is large. 
Perhaps the best example of such a system is the Wolf-Rayet binary 
WR~140 (WC7 + O4V; P = 2900~d;
Williams \etal \cite{W1997}). Luo \etal (\cite{LMM1990}) argue that for 
a WR+O system in the adiabatic limit ``the shocked WR stellar wind dominates
the X-ray emission.''  Myasnikov \& Zhekov (\cite{MZ1993}) arrive at a 
similar conclusion for their ``standard system'' which is also near this
limit: ``the whole of the luminosity is due entirely to the emission of the
shocked gas of the WR wind'', being more than an order of magnitude 
greater than the emission from the weaker wind.\footnote{Although 
Myasnikov \& Zhekov (\cite{MZ1993}) actually use different abundances
for each wind, gas with typical WN abundances has a similar emissivity
as gas with solar abundances, such that this difference
is unimportant to their findings.} 

While the appropriate equations in Usov (\cite{U1992}) are known to be lower
limits (due to the omission of line emission), the ratio of $L_{1}/L_{2}$ 
should not be overly affected and can once again be calculated. Using the 
same values for the stellar wind parameters as Myasnikov \& Zhekov 
(\cite{MZ1993}), Eqs.~89 and~95 in Usov (\cite{U1992}) yield 
$L_{1}/L_{2} = 1.36$, indicating that neither wind is particularly 
dominant. This is clearly in disagreement with the published statements in
Luo \etal (\cite{LMM1990}) and Myasnikov \& Zhekov (\cite{MZ1993}).

To investigate this issue we have calculated the X-ray 
luminosity from numerical simulations of the WCZ near the adiabatic limit. 
Table~\ref{tab:ad_lim} summarizes our findings for winds with equal, 
spatially invariant pre-shock speeds. The general trend is for 
$L_{1}/L_{2}$ to increase with decreasing $\eta$ (which is
opposite to the radiative limit), and for this ratio to
become very large for small $\eta$. We thus find agreement with the 
work of Luo \etal (\cite{LMM1990}) and Myasnikov \& Zhekov 
(\cite{MZ1993}), and conclude that the appropriate equations in 
Usov (\cite{U1992}) are again somewhat in error.

The underlying reason for the trend shown
in Table~\ref{tab:ad_lim} concerns the ratio of the cooling 
timescale to the flow timescale in each of the winds. Stevens \etal 
(\cite{SBP1992}) noted that, near the local minimum in the cooling curve
(which implies that the pre-shock velocity at the stagnation point is in the
range $900 \kmps \ltsimm v \ltsimm 3500 \kmps$), 
this ratio can be approximated as $\chi \approx v_{8}^{4}\;d_{12}/
\Mdot_{-7}$, where $v_{8}$ is the wind velocity in units of 1000\kmps,
$d_{12}$ is the distance from the star to the contact discontinuity in 
units of $10^{12}\;\cm$, and $\Mdot_{-7}$ is the mass-loss rate of the 
star in units of $10^{-7} \Msolpyr$. The shocked wind is radiative 
when $\chi < 1$, and approaches the adiabatic limit when $\chi \gg 1$.
If we ignore the dependence of $\chi$ on $d$ (\ie if we were to assume that 
the relevant distance appropriate to the flow dynamics is the same for 
each wind), the ratio of this characteristic cooling parameter for the two 
winds is $\chi_{1}/\chi_{2} \sim \Mdot_{2} v_{1}^{4}/
\Mdot_{1} v_{2}^{4}$.

For the results in Table~\ref{tab:ad_lim} where $v_{1} = v_{2}$,
$\chi_{1}/\chi_{2} \sim \Mdot_{2}/\Mdot_{1} = \eta$. Hence as the value of 
$\eta$ decreases in Table~\ref{tab:ad_lim}, the stronger wind becomes 
more efficient at emitting X-rays relative to the weaker wind. This is
consistent with the fact that although the post-shock density
at the stagnation point is the same for both winds, its decline with off-axis
distance is faster for the weaker wind (as observed also by Myasnikov 
\& Zhekov \cite{MZ1993}). We note also that within a given distance from 
the stagnation point, the volume occupied by the shocked stronger wind exceeds 
that occupied by the shocked weaker wind.

One might expect that $L_{1}/L_{2} \sim 1$ when $\chi_{1}/\chi_{2} = 1$,
irrespective of the value of $\eta$. Again ignoring the dependence of 
$\chi$ on $d$, we find that $\eta = (v_{2}/v_{1})^5$ for 
$\chi_{1}/\chi_{2} = 1$. Thus to obtain comparable luminosity from each
shocked wind when $\eta = 0.1$, we require $v_{1} \sim 1.6\;v_{2}$. Since,
in reality, mass-loss rates from early-type stars can vary by
several orders of magnitude, whereas wind velocities lie typically within 
the range $1000-3000\;\kmps$ (excluding LBV's), we expect most systems 
near the adiabatic limit will have a luminosity dominated by the stronger wind.

\begin{table}
\begin{center}
\caption{Ratio of the wind X-ray luminosities as a function
of wind momentum ratio, $\eta$, for colliding wind systems with
equal, spatially invariant wind speeds, equal abundances, and near to the 
adiabatic limit.}
\label{tab:ad_lim}
\begin{tabular}{cc}
\hline
$\eta$ & $L_{1}/L_{2}$ \\
\hline
1.0000 & 1.0 \\
0.3160 & 1.9 \\
0.1000 & 3.9 \\
0.0316 & 9 \\
0.0100 & 24 \\
\hline
\end{tabular}
\end{center}
\end{table}

\subsection{$L_{1}/L_{2}$ in-between these limits}
\label{sec:inbetween}

Estimates of the luminosity ratio for systems where the shocked winds are 
in-between the limiting radiative and adiabatic cases must be done on 
a case-by-case basis. Somewhere in this region of parameter space, the wind 
which dominates the  X-ray emission must switch over from the one with the 
faster pre-shock speed (radiative limit) to the one with the slower pre-shock 
speed (adiabatic limit). In contrast, situations where one wind is clearly 
radiative and the other is closer to being adiabatic will have their X-ray 
emission dominated by the former (\eg in $\gamma^{2}$~Velorum, where 
$\chi_{1} \ll 1$ and $\chi_{2} > 1$, the X-ray emission is dominated by the 
shocked WR wind).

\section{Summary}

In this letter we have re-examined the issue of which wind is the dominant
X-ray emitter in colliding wind binaries, following the discovery of some 
confusion in the existing literature. Our work supports the earlier 
conclusions in Luo \etal (\cite{LMM1990}) and Myasnikov \& Zhekov 
(\cite{MZ1993}), though is sometimes in disagreement with the 
relevant analytical equations in Usov (\cite{U1992}). 

For systems near the radiative limit (typically short period binaries), we 
find that the primary influence on $L_{1}/L_{2}$ is the ratio of the wind 
speeds, $v_{1}/v_{2}$, since it controls the energy flux ratio. 
For $v_{1} = v_{2}$, $L_{1} \approx L_{2}$ irrespective of the wind momentum
ratio, $\eta$. When $v_{1} \neq v_{2}$, the faster wind normally dominates
the luminosity, although it is unlikely to do so by more than
a factor of 5. The equations in Usov (\cite{U1992}) predict values for 
$L_{1}/L_{2}$ which can be in error by up to a factor of 4. 

For systems near the adiabatic limit (\eg long-period, high eccentricity 
binaries at apastron), we confirm earlier findings that the stronger 
wind is typically the dominant X-ray emitter, often by an order of 
magnitude relative to the weaker wind. This is because the dominant driver of 
$L_{1}/L_{2}$ is the ratio of $\chi_{1}/\chi_{2}$, so that the stronger wind
can in general more easily radiatively cool. If we force 
$\chi_{1}/\chi_{2} \sim 1$, we find that both shocked winds contribute 
roughly equally to the X-ray emission irrespective of the value of $\eta$. 
In contrast, the equations in Usov (\cite{U1992}) 
yield $L_{1} \sim L_{2}$ irrespective of the assumed
wind parameters.

For systems in-between these limits, we anticipate that the dominant 
luminosity should generally switch from the faster wind (radiative limit) 
to the slower wind (adiabatic limit). Where one shocked 
wind is substantially closer to the radiative limit than that of the 
other, the X-ray emission will naturally be dominated by the former.

This work provides a basic understanding of the dominant factors
controlling the luminosity ratio in colliding wind binaries. Detailed 
calculations are needed to further investigate the effects of wind 
acceleration/braking and different wind abundances. 

\begin{acknowledgements}
We would like to thank Ken Gayley, whose comments led to this work 
and who also as referee helped to clarify some of the important points.
JMP would also like to thank PPARC for the funding of a PDRA position.
\end{acknowledgements}

\end{document}